\documentclass[aps,twocolumn]{revtex4}
\usepackage{graphicx}
\usepackage{amsmath}

\newcommand{\grl}{Geophys. Res. Lett.}

\begin{document}
\title{Nonlinear Whistlerons\footnote{Proceedings of the \textit{International Conference on
Plasma Physics - ICPP 2004}, Nice (France), 25 - 29 Oct. 2004;
contribution P1-019; Electronic proceedings available online at:
\texttt{http://hal.ccsd.cnrs.fr/ccsd-00001894/en/}  .}}
\author{B. Eliasson, I. Kourakis and P. K. Shukla}
\affiliation{Institut f\" ur Theoretische Physik IV, Fakult\"at
f\"ur Physik und Astronomie, Ruhr--Universit\"at Bochum, D--44780
Bochum, Germany}
\date{Received 12 October 2004}
\begin{abstract}
Recently, observations from laboratory experiments have revealed
amplitude modulation of whistlers by low-frequency
perturbations. We here present theoretical and
simulation studies of amplitude modulated whistler solitary waves (whistlerons)
and their interaction with background low-frequency density
perturbations created by the whistler ponderomotive
force. We derive a nonlinear a nonlinear Schr\"odinger equation
which governs the evolution of
whistlers in the presence of finite-amplitude density perturbations, 
and a set of equations for arbitrary large amplitude
density perturbations in the presence of the whistler
ponderomotive force. The governing equations studied analytically in the
small amplitude limit, and are solved numerically to show the existence of large scale density
perturbations that are self-consistently created by localized
whistlerons.  Our numerical results are
in good agreement with recent experimental results where
the the formation of modulated whistlers and solitary whister waves were formed.
\end{abstract}

\maketitle

%
\section{Introduction}
Almost three decades ago, Stenzel \cite{r1} experimentally
demonstrated the creation of a magnetic field-aligned density
cavities by the ponderomotive force of localized electron whistlers.
Observations from a recent laboratory experiment \cite{r2} exhibit
the creation of modulated whistler wavepackets due to
nonlinear effects. Furthermore, instruments on board the CLUSTER
spacecraft have been observing broadband intense electromagnetic
waves, correlated density fluctuations and solitary waves near the
Earth's plasmapause, magnetopause and 
foreshock \cite{r3}, revealing signatures of whistler
turbulence in the presence of density depletions and enhancements.
The Freja satellite \cite{r4} also observed the
formation of envelope whistler solitary waves correlated with
density cavities in the plasma.

A theoretical investigation has in the past predicted the self-channeling
of electron whistlers and the creation of a localized density hump
\cite{r5}. Taking into account the
spatio-temporal dependent whistler ponderomotive force
\cite{r6,r7}, investigations of the modulation and
filamentation of finite amplitude whistlers interacting with
magnetosonic waves \cite{r8,r9,r10} and
ion-acoustic perturbations \cite{r11,r12} have been carried out.

In this article, we investigate nonlinearly interacting
electron whistlers and arbitrary large amplitude ion-acoustic perturbations,
by using computer simulations, and we find analytical expressions for whistlerons 
in the low-amplitude limit \cite{Eliasson,Kourakis}.
\section{Derivation of the governing equations}
Let us consider the propagation of nonlinearly coupled whistlers
and ion-acoustic perturbations in a fully ionized electron-ion plasma
in  a uniform external magnetic field $\widehat {\bf z} B_0$, where
$\widehat {\bf z}$ is the unit vector along the $z$ direction and $B_0$
is the magnitude of the magnetic field strength. We consider the propagation
of right-hand circularly polarized modulated whistlers of the form
\begin{equation}
{\bf E}=\frac{1}{2}E(z,t)(\widehat{\bf{x}}+i\widehat{\bf{y}})
\exp[i(k_0 z-\omega_0 t)]+ \mbox{c.c},
\end{equation}
where $E(z,t)$ is the slowly varying envelope of the whistler
electric field, and $\widehat {\bf x}$ and $\widehat {\bf y}$ are the unit
vectors along the $x$ and $y$ axes, respectively, and $\mbox{c.c.}$ stands
for the complex conjugate. The whistler frequency $\omega_0$ $(\gg
\sqrt{ \omega_{ce}\omega_{ci}})$, and the wavenumber $k_0$ are related
by the cold plasma dispersion relation 
\begin{equation}
  \omega_0=\frac{k_0^2c^2\omega_{ce}}{\omega_{pe,0}^2+k_0^2 c^2},
\end{equation}
where $c$ is the speed of light in vacuum, $\omega_{ce}=eB_0/m_ec$ ($\omega_{ci}
=eB_0/m_ic$) is the electron (ion) gyrofrequency, $\omega_{pe,0}= (4\pi n_0 e^2/m_e)^{1/2}$
is the electron plasma frequency, $e$ is the magnitude of the electron charge,
$m_e$ $(m_i)$ is the electron mass, and $n_0$ is the unperturbed background electron
number density.

The dynamics of modulated whistler wavepacket in the presence
of electron density perturbations associated with low-frequency
ion-acoustic fluctuations and of the nonlinear frequency-shift caused
by the magnetic field-aligned free streaming of electrons (with
the flow speed $v_{ez}$), is governed by the nonlinear Schr{\"o}dinger
equation \cite{r12}
\begin{equation}
  i(\partial_t+v_g\partial_z)E+\frac{v_g'}{2}\partial^2_{zz}E+
  (\omega_0-\omega)E=0,
\end{equation}
where 
\begin{equation}
  \omega=\frac{k_0^2 c^2 \omega_{ce}}{\omega_{pe}^2+k_0^2 c^2} +
  k_0 v_{ez}, 
\end{equation}
and $\omega_{pe}^2=\omega_{pe,0}^2 n_e/n_0$ is the
local plasma frequency including the electron density $n_e$ of the
plasma slow motion. The group velocity and the group dispersion of
whistlers are 
\begin{align}
& v_g=\frac{\partial \omega_0}{\partial k_0}
=2\left(1-\frac{\omega_0}{\omega_{ce}}\right)\frac{\omega_0}{k_0}
\intertext{and}
&v_g'=\frac{\partial^2\omega_0}{\partial k_0^2}=
2\left(1-\frac{\omega_0}{\omega_{ce}}\right)
\left(1-4\frac{\omega_0}{\omega_{ce}}\right)\frac{\omega_0}{k_0^2}, 
\end{align}
respectively.

The equations for the ion motion involved in the low-frequency (in
comparison with the whistler wave frequency) ion-acoustic
perturbations are
\begin{align}
 &\partial_t n_i+\partial_z(n_i v_{iz})=0
\intertext{and}
 &\partial_t v_{iz}+\frac{1}{2}\partial_z v_{iz}^2=
  -\frac{e}{m_i}\partial_z\phi-\frac{1}{m_i n_i}\partial_z p_i,
\end{align}
where, for an adiabatic compression in one space dimension,
the ion pressure is given by $p_i=p_{i,0}(n_i/n_0)^{3}$.
Here, the unperturbed ion pressure is denoted
by $p_{i,0}=n_0 T_i$, where $T_i$ is the ion temperature.

The electron dynamics in  the plasma slow motion is
governed by the continuity and momentum equations, {\it viz.}
\begin{align}
  &\partial_t n_e+\partial_z(n_e v_{ez})=0
\intertext{and}
  &0=\frac{e}{T_e}\partial_z\phi-\partial_z
  \mathrm{ln}\left(\frac{n_e}{n_0}\right)+ F,
\end{align}
where $T_e$ is the electron temperature, $\phi$ is the ambipolar potential,
and the low-frequency ponderomotive force of electron whistlers is
\begin{equation}
   F=\frac{\omega_{pe,0}^2}{\omega_0 (\omega_{ce}-\omega_0)}
   \left(\partial_z+\frac{2}{v_g}\partial_t\right)
   \frac{|E|^2}{4\pi n_0 T_e}.
\end{equation}
The system of equations is closed by means of
quasi--neutrality 
\begin{equation}
  n_i= n_e\equiv n, 
\end{equation}
which is justified if
$\omega_0 <\omega_{ce}$ is fulfilled with some margin. 
The continuity equations for the electrons and ions give $v_{iz}=v_{ez}\equiv
v_z$, so that 
\begin{equation}
  \partial_t n+\partial_z(n v_{z})=0.
\end{equation}
Eliminating $\partial_z\phi$ from the governing equations for low-
frequency density perturbations, we have
\begin{equation}
   \partial_t v_{z}+\frac{1}{2}\partial_z v_{z}^2=
  -\frac{T_e}{m_i}\left[\partial_z \mathrm{ln}\left(\frac{n}{n_0}\right)-F\right]
  -\frac{1}{m_i n}\partial_z p_i.
\end{equation}
The nonlinear Schr\"odinger equation for the whistler electric field
together with the low-frequency equations form a closed set for our
purposes.
\subsection{Dimensionless variables}
In order to investigate numerically the interaction between
whistlers and large amplitude ion-acoustic perturbations, it is
convenient to normalize the governing equations into dimensionless
units, so that relevant parameters can be chosen. We introduce
the dimensionless variables $\xi=\omega_{pi,0}z/C_s$, where the
sound speed is $C_s=[(T_e+3T_i)/m_i]^{1/2}$,
$\tau=\omega_{pi,0}t$, $N=n/n_0$, $u=v_z/C_s$ and ${\cal
E}=E/\sqrt{4\pi n_0 (T_e+3T_i)}$; the only free dimensionless
parameters of the system are
$\Omega_{c}=\omega_{ce}/\omega_{pi,0}$, $\kappa=k_0
c/\omega_{pe,0}$, $\eta=T_i/T_e$ and $V_g=v_g/C_s$. The normalized system of
equations are of the form
\begin{align}
  &\partial_{\tau}N=-\partial_{\xi}(Nu),
  \\
  \begin{split}
  &\partial_{\tau}\left(u-\frac{2\alpha}{V_g} |{\cal E}|^2\right)
  \\
  &=\partial_{\xi}\left[-\frac{u^2}{2}-
  \frac{\mathrm{ln} N+1.5\eta N^2}{1+3\eta}
  +\alpha |{\cal E}|^2\right],
  \end{split}
  \intertext{and}
\begin{split}
  &\partial_{\tau} {\cal E}=-V_g \partial_{\xi} {\cal E}
  +i\bigg[P\partial^2_{\xi\xi} {\cal E}
  \\
  &+\left(\frac{1}{1+\kappa^2}-\frac{1}{N+\kappa^2}
  -\frac{2}{(1+\kappa^2)^2}\frac{u}{V_g}\right)
  \Omega_{c} \kappa^2 {\cal E}\bigg],
\end{split}
\end{align}
where the constants are
$\alpha=(1+\kappa^2)^2\omega_{pe,0}^2/\omega_{ce}^2\kappa^2$ and
$P=(1+\kappa^2)(1-3\kappa^2)V_g^2/4\kappa^2\Omega_c$. The sign of
the coefficient $P$, multiplying the dispersive term in  Eq. (3),
depends on $\kappa$: When $\kappa<1/\sqrt{3}$, $P$ is positive and
for $\kappa<1/\sqrt{3}$ we see that $P$ is negative.
\section{Small-amplitude solitary waves}
In the small-amplitude limit, viz. $N=1+N_1$, $u=u_1$, where
$N_1$, $u_1 \ll 1$, Eqs. (1)--(3) yield
\begin{align}
  &\partial_{\tau}N_1=-\partial_{\xi}u_1,
  \\
  &\partial_{\tau}\left(u_1-\frac{2\alpha}{V_g} |{\cal E}|^2\right)=
  \partial_{\xi}\left(-N_1
  +\alpha |{\cal E}|^2\right),
  \intertext{and}
\begin{split}
  &\partial_{\tau} {\cal E}=-V_g \partial_{\xi} {\cal E}
\\
  &+i\left[P\partial^2_{\xi\xi} {\cal E}
  +\left(N_1-\frac{u}{V_g}\right)
  \frac{\Omega_{c} \kappa^2{\cal E}}{(1+\kappa^2)^2}
  \right],
\end{split}
\end{align}
where the only nonlinearity kept is the ponderomotive force terms
involving $|{\cal E}|^2$.
It is important to remember that our nonlinear Schr\"odinger equation
for the whistler field is based on a Taylor expansion of the dispersion
relation for whistler waves around a wavenumber $k_0$.
Thus, this model is only accurate for
wave envelopes moving with speeds close to the group speed $V_g$, and
other speeds of the wave envelopes may give unphysical results.
Here, we look for whistler envelope solitary
waves moving with the group speed $V_g$, so that $N_1$ and $u_1$ depends
only on $\chi=\xi-V_g\tau$, while the electric field envelope is
assumed to be of the form ${\cal E}=W(\chi)\exp(i\Omega\tau-ik\xi)$,
where $W$ is a real-valued function of one argument. Using the boundary
conditions $N_1=0$, $u_1=0$ and $W=0$ at $|\xi|=\infty$, we have
$k=0$, $N_1=-W^2 \alpha/(1-V_g^2)$ and $u_1=V_g N_1$. We here note that
subsonic ($V_g<1$) solitary waves are characterized by a density cavity
while supersonic ($V_g>1$) envelope solitary waves are characterized by a
density hump. The system of equations (4) to (6) is then reduced to
the cubic Schr\"odinger equation
\begin{equation}
  P\partial^2_{\chi\chi}W+QW^3-\Omega W=0,
\end{equation}
where $Q=\alpha\Omega_c\kappa^2/(1+\kappa^2)(1-V_g^2)$. Localized
solutions of Eq. (7) only exist if the product $PQ$ is positive.
We note that $P>0$ ($P<0$) when the whistler frequency
$\omega_0<\omega_{ce}/4$ ($\omega_0>\omega_{ce}/4$), and that
$Q<0$ ($Q>0$) when $|V_g|<1$ ($|V_g|>1$), so in the frequency band
where $\omega_0<\omega_{ce}/4$, only subsonic solitary waves,
characterized by a localized density cavity can exist, while in
the frequency band $\omega_0>\omega_{ce}/4$, only supersonic
solitary waves characterized by a localized density hump exist.
Equation (7) has exact solitary wave solutions of the form
\begin{equation}
  W=\left(\frac{2\Omega}{Q}\right)^{1/2}
  \mathrm{sech}\left[\left(\frac{\Omega}{P}\right)^{1/2}(\xi-V_g\tau-\xi_0)\right],
\end{equation}
where $V_g$ and $\Omega$ and the displacement $\xi_0$ are the
three free parameters for a given set of physical plasma
parameters. Finally, we recall that the dispersion relation for the electron
whistlers used here is valid if $\omega_0>\sqrt{\omega_{ce}\omega_{ci}}$.
For subsonic whistlers having the group speed $v_g=C_s V_g$ (where $V_g<1$),
where $v_g\approx 2\omega_0/k_0$ and $\omega_0\approx k_0^2c^2\omega_{ce}/\omega_{pe,0}^2$,
we have  $ck_0/\omega_{pe,0}=(C_s/c)(\omega_{pe,0}/\omega_{ce})V_g/2>(m_e/m_i)^{1/4}$.
\section{Numerical results}
We have investigated the properties of modulated whistler wave packets by
solving numerically Eqs. (1)--(3).  We have here chosen parameters
from a recent experiment, where the formation of localized
whistler envelopes have been observed \cite{r2}.
In the experiment, one has $n_0=1.2\times 10^{12}\,\mathrm{cm}^{-3}$ and $B_0=100$ G,
so that $\omega_{pe,0}=6.7\times10^{10}\,\mathrm{s}^{-1}$ and
$\omega_{ce}=1.76\times10^9\,\mathrm{s}^{-1}$,
respectively. Hence, $\omega_{ce}/\omega_{pe,0}=0.026$. The frequency of the whistler wave
is $\omega_0=2\pi\times 160\times10^6\,\mathrm{s}^{-1}=1.0\times10^9\,s^{-1}$, so that
$\omega_0/\omega_{ce}\approx 0.57>0.25$. Thus, the whistlers have negative group dispersion.
From the dispersion relation of whistlers, we have $\kappa\approx 1.15$, which gives
$k_0\approx 257\,\mathrm{m}^{-1}$. The latter corresponds to whistlers with a wavelength
of 2.4 cm. Furthermore, the whistler group velocity is $v_g=3.36\times10^6$ m/s.
The argon ion-electron plasma ($m_i/m_e=73400$) had the temperatures
of $T_e=10$ eV and $T_i=0.5$ eV, giving the sound speed $5.25\times10^3\,\mathrm{m/s}$, and
the normalized group velocity $V_g=v_g/C_s=640$.
\begin{figure}[floatfix]
\includegraphics[width=\columnwidth]{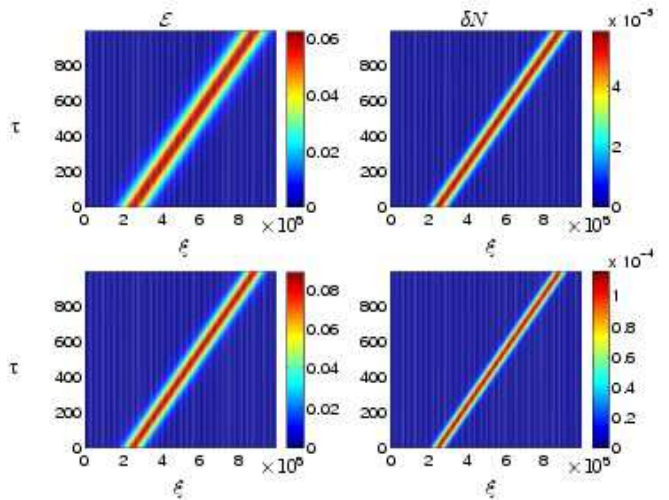}
\bigskip
\caption{
Simulations of solitary whistler waves, with its
associated electric field envelope $|{\cal E}|$ (left panels)
and density perturbation $\delta N=N-1$ (right panels).
Parameters are: $V_g=640$, $\kappa=1.15$,
$\Omega_c=\omega_{ce}/\omega_{pi,0}=7.05$, $m_i/m_e=73400$ (argon ions) and
$\eta=0.05$. For the initial condition, we used the small-amplitude solitary wave
solution ${\cal E}=W=(2\Omega/Q)^{1/2}\mathrm{sech}[(\Omega/P)^{1/2}(\xi-2.5\times10^5)]$,
with $Q=-0.025$ and $P=-7.57\times10^4$ for the given parameters,
and $\Omega=-0.5\times10^{-4}$ (upper panels) and
$\Omega=-1\times10^{-4}$ (lower panels). For the density and
velocity, we used $N=1+N_1$ and $u=u_1$, where $N_1=-W^2\alpha/(1-V_g^2)$ and $u_1=V_g N_1$.
}
\end{figure}
In Fig. 1, we have illustrated localized whistler envelope
solitons, in which the electric field envelope (left panels) is accompanied
with a density hump (right panels). We notice that the density hump is
relatively small, due to the large (in comparison with the acoustic speed) 
group velocity of the whistler waves.
\begin{figure}[floatfix]
\includegraphics[width=\columnwidth]{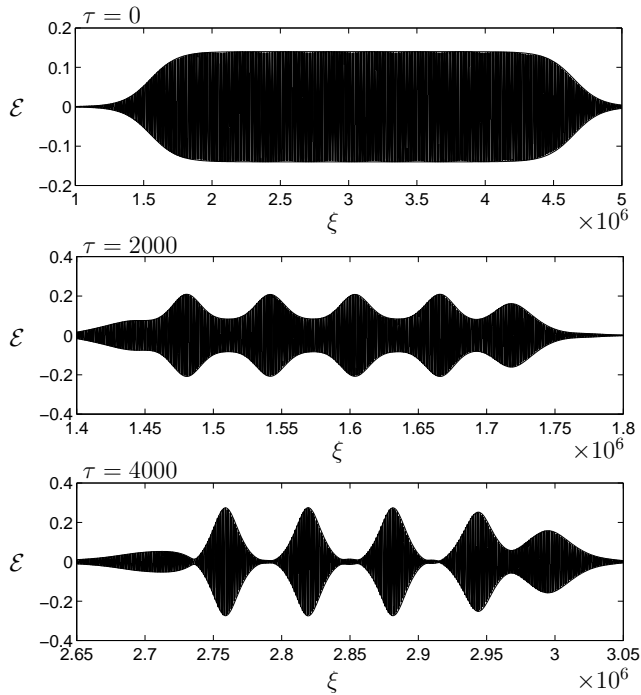}
\bigskip
\caption{The electric field
as a function of normalized space
$\xi$ at the times $\tau=0$ (upper panel),
$\tau=2000$ (middle panel) and $\tau=4000$ (lower panel).
The initial condition was a pulse on the form
${\cal E}=0.07\{1-\tanh[5\cos(\pi \xi/3.1\times 10^5)]\}$,
while the density was perturbed as
$N=1+0.01\cos(20\pi\xi/6.2\times10^5)$. Parameters are: $V_g=640$, $\kappa=1.15$,
$\Omega_c=\omega_{ce}/\omega_{pi,0}=7.05$, $m_i/m_e=73400$, $\eta=0.05$.
The (scaled by $C_s/\omega_{pi,0}$) wavelength of the high-frequency
wave is $\lambda_0=1.16\times10^3$,
corresponding to $\sim2.5$ cm in dimensional units. The envelope
$\pm |{\cal E}|$ was plotted together with the real part of the
wave ${\cal E}\exp(i K_0\xi)$, where $K_0=k_0 C_s/\omega_{pi}=2\kappa^2\Omega_c/(1+\kappa^2)^2V_g$.
We see the formation of separate wavepackets
of the modulated high-frequency wave.
}
\end{figure}
In Fig. 2, we can see the development of a large-amplitude
whistler pulse, which has been launched in a plasma perturbed by
ion-acoustic waves, with a density modulation of one percent
(see the caption of Fig. 2). This simulates, to some extent, the experiment
by Kostrov {\it et al.}, where the density and magnetic field were perturbed
by a low-frequency conical refraction wave, giving rise to a modulation
of the electron whistlers. Here, as in the experiment,
we observe that a modulated electron whistler pulse (middle panel of Fig. 2) develops
into isolated solitary electron whistler waves (lower panel).  We note that the wavelength
of the whistlers is $\approx 2.5$ cm, while the typical width
of a solitary pulse is $\Delta\xi\approx 3\times10^4$ in the scaled length
units, corresponding to $\approx 64$ cm, so that each solitary wave train contains
25 wavelengths of the high-frequency whistlers. In one experiment,
illustrated in the lower panel of Fig 4 in
Ref. \cite{r2},
one finds that the width of the solitary whistler pulse in time is $0.2\,\mathrm{\mu s}$,
which with the group speed $v_g=3.36\times 10^6$ m/s gives the width $\sim 60$ cm
in space of the solitary wave packets, in good agreement with our numerical results.
From the relation $N_1=-W^2\alpha/(1-V_g^2)$ valid for solitary whistlers in
the small-amplitude limit, and with the amplitude of $W=|{\cal E}|$ approximately
$0.3$ seen in the lower panel of Fig. 2, we can estimate the relative amplitude of the
density hump associated with the solitary waves to be of the order $10^{-3}$,
i.e. much smaller than the modulation $\sim 10^{-2}$ due to the ion-acoustic
waves excited in the initial condition.

Next, we study the properties of subsonic whistler envelope solitary pulses
which have the normalized group speed $V_g=0.5$.  Here, the restrictive condition
$ck_0/\omega_{pe,0}=(C_s/c)(\omega_{pe,0} /\omega_{ce})V_g/2>(m_e/m_i)^{1/4}$
requires somewhat higher values of the plasma temperature and $\omega_{pe,0}/\omega_{ce}$
for their existence. With $m_i/m_e=30000$, we have $(m_e/m_e)^{1/4}\approx 0.1$. We
take $\kappa=ck_0/\omega_{pe,0}=0.2$, $C_s=10^5$ m/s (corresponding to $T_e\sim 1400$ eV)
$\eta=0.1$, and $\omega_{pe,0}/\omega_{ce}=2400$. Thus, $\Omega_c=0.072$ and
$\omega_0/\omega_{ce}\approx 0.039$. For these values of the parameters, there exist
solitary whistler pulse  solutions, which we have displayed in Fig. 3.
\begin{figure}[floatfix]
\includegraphics[width=\columnwidth]{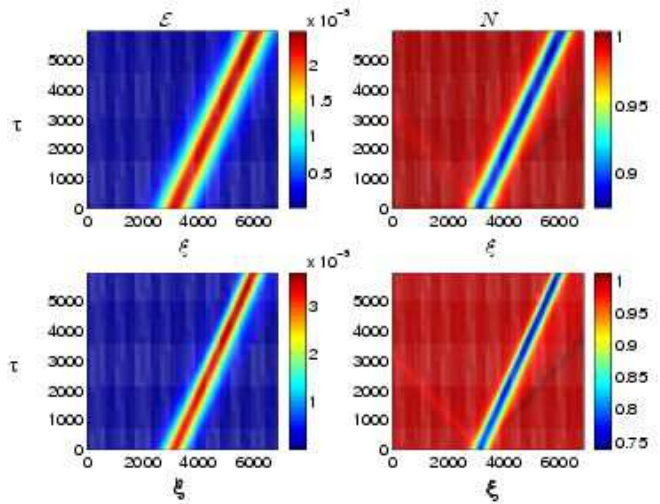}
\bigskip
\caption{Simulations of solitary whistler waves, with its
associated electric field envelope $|{\cal E}|$ (left panels)
and density $N$ (right panels). Parameters are: $V_g=0.5$, $\kappa=0.2$,
$\Omega_c=\omega_{ce}/\omega_{pi,0}=0.072$, $m_i/m_e=30000$, $\eta=0.1$.
For the initial condition, we used the small-amplitude solitary wave
solution ${\cal E}=W=(2\Omega/Q)^{1/2}\mathrm{sech}[(\Omega/P)^{1/2}(\xi-3000)]$,
with $Q=5.5\times10^5$ and $P=10.8$ for the given parameters,
and $\Omega=1.5\times10^{-4}$ (upper panels) and
$\Omega=3\times10^{-4}$ (lower panels). For the density and
velocity, we used $N=1+N_1$ and $u=u_1$, where
$N_1=-W^2\alpha/(1-V_g^2)$ and $u_1=V_g N_1$.
}
\end{figure}
We have used the exact solution in the small-amplitude limit
as an initial condition for the simulation of the full system of equations (1)--(3).
The bell-shaped whistler electric field envelope is accompanied with a large-amplitude
plasma density cavity.
\section{Discussion}
We have presented theoretical and simulation
studies of nonlinearly interacting electron whistlers and
arbitrary large amplitude ion-acoustic perturbations in a
magnetized plasma. For this purpose, we have derived a set
of equations which describe the spatio-temporal evolution of a
modulated whistler packet in the present of slowly varying plasma
density perturbations. The ponderomotive force of the latter,
in turn, modifies the local plasma density in a self-consistent
manner. Numerical solutions of the governing nonlinear equations reveal
that subsonic envelope whistler solitons are characterized by a bell-
shaped whistler electric fields that are trapped in self-created
density cavity. This happens when the  whistler wave frequency is
smaller than $\omega_{ce}/4$, where the waves have positive group
dispersion. When the whistler wave frequency is larger than
$\omega_{ce}/4$, one encounters negative group dispersive
whistlers and the supersonic whistler envelope solitons are
characterized by a bell-shaped whistler electric fields which create
a density hump. Modulated whistler wavepackets have indeed been observed
in a laboratory experiment \cite{r2} as
well as near the plasmapause \cite{r3}
and in the auroral zone \cite{r4}.
Our results are in excellent agreement with the experimental results
\cite{r2}, while we think that a multi-dimensional study, including channelling of
whistler waves in density ducts, is required to interpret the observations
by Cluster and Freja satellites.
\acknowledgments
This work was partially supported by the European Commission (Brussels,
Belgium) through contract No. HPRN-CT-2001-00314, as well as by the
Deutsche Forschungsgemeinschaft through the Sonderforschungsbereich
591.


\begin{thebibliography}{99}

\bibitem{r1} 
Stenzel R. L., Filamentation of large amplitude
whistler waves, \grl {\bf 3}, 61-64 (1976).

\bibitem{r2}
Kostrov A. V., Gushchin, M. E., Korobkov, S. V., and
Strikovskii A. V., Parametric Transformation of the amplitude and
frequency of a whistler wave in a magnetoactive plasma, {\it JETP
Lett.} {\bf 78}, 538-541 (2003).

\bibitem{r3}
Moullard O., Masson A., Laasko H. {\it et al.}, Density modulated
whistler mode emissions observed near the plasmapause, \grl {\bf
29}, doi:10.1029/2002GL015101 (2002).

\bibitem{r4}
Huang G. L., Wang D. Y., and Song, Q. W., Whistler waves in Freja
observations, {\it J. Geophys. Res.} {\bf 109}, A02307,
doi:10.1029/2003JA011137 (2004).

\bibitem{r5}
Weibel E. S., Self-channeling of whistler waves,
{\it Phys. Lett.} {\bf 61A}, 37-39 (1977).

\bibitem{r6}
Washimi H. and Karpman V. I., The ponderomotive force of a
high-frequency electromagnetic field in a dispersive medium, {\it
Soviet Phys. JETP} {\bf 44}, 528-531 (1976).

\bibitem{r7}
Tskhakaya D. D. , On the `non-stationary' ponderomotive force of a
HF field in a plasma, {\it J. Plasma Phys.} {\bf 25}, 233-239 (1981).

\bibitem{r8}
Hasegawa A., Stimulated modulational instabilities of plasma waves,
{\it Phys. Rev. A} {\bf 1}, 1746-1750 (1970).

\bibitem{r9}
Karpman V. I. and Washimi, H.,  Two-dimensional self-modulation of a
whistler wave propagating along the magnetic field in a plasma,
{\it J. Plasma Phys.} {\bf 18}, 173-187 (1977).

\bibitem{r10}
Karpman V. I. and Stenflo L., Equations describing the interaction
between whistlers and magnetosonic waves, {\it Phys. Lett. A} {\bf 127},
99-101 (1988).

\bibitem{r11}
Bogolybskii I. L. and Makha'nkov V. G., Energy-conversion mechanism in
the formation and interaction of helicon solitons, {\it Sov.
Phys. Tech. Phys.} {\bf 21}, 255-258 (1976).

\bibitem{r12}
Spatschek, K. H., Shukla P. K., Yu M. Y. {\it et al.}, Finite
amplitude localized whistler waves, {\it Phys. Fluids} {\bf 22},
576-582 (1979).

\bibitem{Eliasson} B. Eliasson and P. K. Shukla, Theoretical and numerical
study of density modulated whistlers, {\it Geophys. Res. Lett.} {\bf 31}, L17802,
doi:10.1029/2004GL020605 (2004).

\bibitem{Kourakis} I. Kourakis and P. K. Shukla, Modulated whistler wavepackets
associated with density perturbations, {\it Phys. Plasmas} (in press 2004).
\end{thebibliography}
\end{document}